\documentclass[10pt,twocolumn]{IEEEtran}
\usepackage{graphicx,epsfig}
\usepackage{latexsym,subfigure,amsmath}
\usepackage{cite}
\begin{document}

\title{Fingerprints in the Ether: Using the Physical Layer for
Wireless Authentication}
\author{Liang Xiao, Larry Greenstein, Narayan Mandayam, Wade Trappe
\thanks{The authors may be reached at \{lxiao, ljg, narayan, trappe\}@winlab.rutgers.edu. This
research is supported, in part, through a grant CNS-0626439 from the National Science Foundation.}\\
Wireless Information Network Laboratory (WINLAB), Rutgers University\\
671 Rt. 1 South, North Brunswick, NJ 08902} \maketitle

\begin{abstract}
The wireless medium contains domain-specific information that can be
used to complement and enhance traditional security mechanisms.  In
this paper we propose ways to exploit the fact that, in a typically
rich scattering environment, the radio channel response decorrelates
quite rapidly in space. Specifically, we describe a physical-layer
algorithm that combines channel probing ($M$ complex frequency
response samples over a bandwidth $W$) with hypothesis testing to
determine whether current and prior communication attempts are made
by the same user (same channel response). In this way, legitimate
users can be reliably authenticated and false users can be reliably
detected. To evaluate the feasibility of our algorithm, we simulate
spatially variable channel responses in real environments using the
WiSE ray-tracing tool; and we analyze the ability of a receiver to
discriminate between transmitters (users) based on their channel
frequency responses in a given office environment. For several rooms
in the extremities of the building we considered, we have confirmed
the efficacy of our approach under static channel conditions. For
example, measuring five frequency response samples over a bandwidth
of $100$ MHz and using a transmit power of $100$ mW, valid users can
be verified with $99$\% confidence while rejecting false users with
greater than $95$\% confidence.
\end{abstract}

\section{Introduction}
As wireless devices become increasingly pervasive and essential,
they are becoming both a target for attack and the very weapon
with which such an attack can be carried out. Traditional
high-level computer and network security techniques can, and must,
play an important role in combating such attacks, but the wireless
environment presents both the means and the opportunity for new
forms of intrusion. The devices that comprise a wireless network
environment are low-cost commodity items that are easily available
to potential intruders and also easily modifiable for such
intrusion. In particular, wireless networks are open to intrusion
from the outside without the need for a physical connection and,
as a result, techniques which would provide a high level of
security in a wired network have proven inadequate in a wireless
network, as many motivated groups of students have readily
demonstrated\cite{Borisov:802.11,Arbaugh:Clothes, Walker:WEP}.

Although conventional cryptographic security mechanisms are
essential to securing wireless networks, these techniques do not
directly leverage the unique properties of the wireless domain to
address security threats. The physical properties of the wireless
medium are a powerful source of domain-specific information that can
be used to complement and enhance traditional security mechanisms.
In this paper, we propose that a cross-layer approach can be used to
augment the security of wireless networks. In particular, we believe
that the nature of the wireless medium can be turned to the
advantage of the network engineer when trying to secure wireless
communications. The enabling factor in our approach is that, in the
rich multipath environment typical of wireless scenarios, the
response of the medium along any transmit-receive path is {\em
frequency-selective} (or in the time domain, {\em dispersive}) in a
way that is {\em location-specific}. This means:
\begin{enumerate}
\item The channel can be specified by a number of complex samples
either in the frequency domain (a set of complex gains at a set of
frequencies) or the time domain (a set of impulse response samples
at a set of time delays). \item Such sets of numbers decorrelate
from one transmit-receive path to another if the paths are separated
by the order of an RF wavelength or more. \end{enumerate} \noindent
Using the uniqueness of the channel between two locations,
we believe it is possible to establish new forms of authentication
that include information available at the physical layer. Rather than rely solely on
higher-layer cryptographic mechanisms, wireless devices can
authenticate themselves based upon their ability to produce an
appropriate signal at the recipient.

%***save this for a punch line***
%These unique space, time, and frequency characteristics of the
%wireless physical layer can be used to augment traditional
%higher-layer authentication. In particular, while traditional
%authentication techniques are robust, they are often applied only
%periodically, leaving systems open to intrusion between moments of
%authentication. Physical layer approaches, however, can fill the gap
%between moments of authentication by providing a continuous means to
%detect an intrusion on the wireless network and can provide warnings
%to initiate more stringent higher-layer security methods.  In short,
%merely using cryptographic methods does not capture the full
%spectrum of possible solutions that are available to the wireless
%engineer.

While using the physical layer to enhance security might seem
to be a radical paradigm shift for wireless systems, we note that
this is not the first time that multipath and advanced physical
layer methods have proven advantageous. Specifically, we are encouraged in
our belief by two notable parallel paradigm shifts in wireless
systems: (1) code division multiple access (CDMA) systems
\cite{Book:CDMAVITERBI}, where the use of Rake processing
transforms multipath into a diversity-enhancing benefit; and (2)
multiple-input multiple-output (MIMO) antenna techniques
\cite{MIMO:original}, which transform scatter-induced Rayleigh
fading into a capacity-enhancing benefit.

In order to support the use of physical layer information for
enhancing wireless security, it is necessary to understand the
degree to which physical layer measurements can serve to
discriminate between transmitters, and then to place this
functionality in the context of a greater end-to-end security
framework. In this paper, we tackle the first of these problems by
providing an initial investigation into the ability of a receiver to
distinguish between transmitters.

We begin the paper in Section \ref{sec:Overview} by providing an
overview of our proposed PHY-layer authentication service. We then
examine the possibilities of achieving physical-layer
authentication using a hypothesis testing framework in Section
~\ref{sec_ana}. In order to validate our ideas, we have performed
simulations using the WiSE propagation tool, and our results are
in Section ~\ref{sec_sim}. Our objective is to understand the
degree to which physical layer authentication is possible, and
hence our initial performance studies reported in this paper are
for a benign, static multipath environment. We wrap up the paper
in Section \ref{sec:Conclusion} by providing concluding remarks
and highlighting important areas for further investigation.

\section{Problem Overview} \label{sec:Overview}

\begin{figure}[t]
\begin{center}
%\begin{tabular}{cc}
%\epsfig{figure=./fig/SEVILLEOverview.eps,width=3.4in} &
       \epsfig{figure=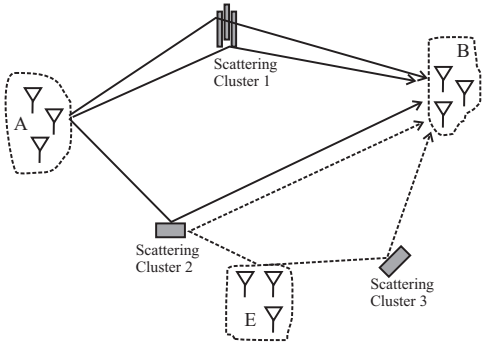,width=2.5in}
 %      (a) & (b)
%\end{tabular}
\end{center}
\vspace{-3mm}\caption{\label{fig:SEVILLE} The adversarial multipath
environment involving multiple scattering surfaces. The transmission
from Alice (A) to Bob (B) experiences different multipath effects
than the transmission by the adversary, Eve (E).}
%\end{wrapfigure}
\end{figure}

Traditionally, authentication involves the verification of an
entity's identity. In the context of physical layer
authentication, however, we are not interested in identity,
\textit{per se}, but rather are interested in recognizing a
particular transmitter device. The ability to distinguish between
different transmitters would be particularly valuable in real
wireless systems, as it would help prevent spoofing attacks, where
one wireless device claims to be another wireless device.
Currently, spoofing attacks are very easy to launch in many
wireless networks. For example, in commodity networks, such as
802.11 networks, it is easy for a device to alter its MAC address
by simply issuing an {\tt ifconfig} command. This weakness is a
serious threat, and there are numerous attacks, ranging from
session hijacking\cite{Arbaugh:Analysis} to attacks on access
control lists\cite{Arbaugh:Clothes}, that are facilitated by the
fact that an adversarial device may masquerade as another device.

Here, we seek to develop the notion of physical-layer
authentication services by making use of the complexity associated
with multipath propagation. Throughout the discussion, we shall
borrow from the conventional terminology of the security community
by introducing three different parties: Alice, Bob and Eve. For
our purposes, these three entities may be be thought of as
wireless transmitters/receivers that are potentially located in
spatially separated positions, as depicted in Figure
\ref{fig:SEVILLE}. Our two ``legal" protagonists are the usual
Alice and Bob, and for the sake of discussion throughout this
paper, Alice will serve as the transmitter that initiates
communication, while Bob will serve as the intended receiver.
Their nefarious adversary, Eve, will serve as an active opponent
who injects undesirable communications into the medium in the
hopes of impersonating Alice.

Our security objective, broadly speaking, is to provide
authentication between Alice and Bob, despite the presence of Eve.
Authentication is traditionally associated with the assurance that a
communication comes from a specific entity\cite{Wadebook}. Returning
to our communication scenario, this objective may be interpreted as
follows. Since there is a potential adversary, Eve, who is within
range of Alice and Bob, and who is capable of injecting her own
signals into the environment to impersonate Alice, it is desirable
for Bob to have the ability to differentiate between legitimate
signals from Alice and illegitimate signals from Eve. He therefore
needs some form of evidence that the signal he receives did, in
fact, come from Alice.

In a multipath environment, the property of rapid spatial
decorrelation can be used to authenticate a transmitter. To
illustrate this, let us return to Figure \ref{fig:SEVILLE} and
consider a simple transmitter identification protocol in which Bob
seeks to verify that Alice is the transmitter. Suppose that Alice
probes the channel sufficiently frequently to assure temporal
coherence between channel estimates and that, prior to Eve's
arrival, Bob has estimated the Alice-Bob channel. Now, Eve wishes to
convince Bob that she is Alice. Bob will require that each
information-carrying transmission be accompanied by an authenticator
signal. The channel and its effect on a transmitted signal between
Alice and Bob is a result of the multipath environment. Bob may use
the received version of the authenticator signal to estimate the
channel response and compare this with a previous record for the
Alice-Bob channel. If the two channel estimates are ``close" to each
other, then Bob will conclude that the source of the message is the
same as the source of the previously sent message. If the channel
estimates are not similar, then Bob should conclude that the source
is likely not Alice.

There are several important issues related to such a procedure
that should be addressed before it can be a viable authentication
mechanism. First is the specification of the authenticator signal
that is used to probe the channel. There are many standardized
techniques to probe the channel, ranging from pulse-style probing
to multi-tonal probing \cite{RappaportBook}, and we may use these
techniques to estimate the channel response. Regardless of what
probing method is employed, the channel response can be
characterized in the frequency domain, and throughout this paper
we will represent our channels in that domain.

Next, at the heart of our idea, we use the fact in a richly
scattered multipath environment (typical of indoor wireless
environments) it is difficult for an adversary to create or
precisely model a waveform that is transmitted and received by
entities that are more than a wavelength away from the adversary.
The difficulty of an adversary to predict the environment is
supported by the well-known Jakes uniform scattering model
\cite{Jakes}, which states that the received signal rapidly
decorrelates over a distance of roughly half a wavelength, and
that spatial separation of one to two wavelengths is sufficient
for assuming independent fading paths. The implication of such a
scattering model in a transmitter identification application
remains to be tested, and one of the objectives behind this study
is to examine the utility of a typical indoor multipath
environment for discriminating between Alice-Bob and Eve-Bob
channels. It should also be noted that the multipath channel will
change with time due to both terminal mobility and changes in the
environment. As mentioned earlier, in practice it will be
necessary to guarantee the continuity of the authentication
procedure by probing the channel at time intervals less than the
channel's coherence time. However, even before issues of temporal
variability can be brought into the picture, it is necessary to
first examine the ability to distinguish between transmitters in a
static multipath environment. This paper examines the ability to
authenticate transmitters in such an environment, and serves to
illustrate the potential for new forms of physical layer security.

\section{Analysis}\label{sec_ana}

In this section, we provide a formulation of physical layer
authentication as a hypothesis testing problem.

\subsection{System Model}
We assume that Bob first measures and stores the frequency response
of the channel connecting Alice with him. Though the true channel
response is $H_{AB}(f)$, Bob stores a noisy version,
$\hat{H}_{AB}(f)$, due to his receiver noise. After a while, he has
to decide whether a transmitting terminal is still Alice, his
decision being based on a noisy measured version, $\hat{H}_t(f)$, of
that terminal's channel response to Bob (the true response being
$H_t(f)$). By sampling $\hat{H}_{AB}(f)$ and $\hat{H}_{t}(f)$, $f\in
(f_o-W/2,f_o+W/2]$, Bob obtains two vectors
$\hat{\underline{H}}_{AB}$ and $\hat{\underline{H}}_{t}$,
\begin{align}
\hat{\underline{H}}_{AB}&=\underline{H}_{AB}e^{j\phi _1}+\underline{N}_1\label{eq_b1}\\
\hat{\underline{H}}_{t}&=\underline{H}_{t}e^{j\phi
_2}+\underline{N}_2\label{eq_b2}
\end{align}
where the elements of vector $\underline{A}=[A_1,\cdots,A_M]^T$ are
samples from $A(f)$. More specifically, $A_m=A(f_o-W/2+m \Delta f)$,
$m=1,\cdots,M$, where $\Delta f=W/M$; $M$ is the sample size; $W$ is
the measurement bandwidth; $f_o$ is the center frequency of the
measurement; and all elements of $\underline{N}_1$ and
$\underline{N}_2$ are i.i.d complex Gaussian noise samples
$CN(0,\sigma ^2)$. Considering the fact that the phase of Bob's
receiver local oscillator (LO) can change between one measurement
and another, we introduce $\phi _1$ and $\phi _2 \in [0,2 \pi)$ to
represent measurement errors in the phase of the channel frequency
response.

\subsection{Hypothesis Testing}
Bob uses a simple hypothesis test \cite{Book:Goodman} to decide if
the transmitting terminal is Alice or a would-be intruder, e.g.,
Eve. The null hypothesis, ${\cal H}_0$, is that the terminal is
not an intruder, i.e. the claimant is Alice; and Bob accepts this
hypothesis if the test statistic he computes, $L$, is below some
threshold, $k$. Otherwise, he accepts the alternative hypothesis,
${\cal H}_1$, that the claimant terminal is someone else.
\begin{align}
{\cal H}_0: &\underline{H}_{t} = \underline{H}_{AB}\label{test1}\\
{\cal H}_1: &\underline{H}_{t} \neq
\underline{H}_{AB}\label{test2}
\end{align}

The test statistic is chosen to be
\begin{align}
L=\min _{\phi}\frac{1}{\sigma ^2} \sum_{m=1}^M \mid
\hat{{H}}_{tm}-\hat{{H}}_{ABm} e^{j \phi} \mid^2 .
\end{align}

The minimization over the phase $\phi$ is necessary to account for
measurement errors in the phase of the frequency response, $\phi _1$
and $\phi _2$. Without this adjustment by Bob, the transmitting
terminal can be rejected even if it is in fact Alice. It is easy to
show that the minimizing value of $\phi$ is
\begin{align}\label{l_phi}
\phi ^* =Arg(\sum_{m=1}^M \hat{H}_{tm} \hat{H}^*_{ABm}) .
\end{align}

For the sake of analytical tractability, we will use for $\phi ^*$
the value corresponding to a noiseless channel ($\hat{H}_{AB}(f) =
H_{AB}(f)$ and $\hat{H}_t(t) = H_t(f)$); for the high-SNR
conditions where the system must operate, this is a very
reasonable approximation.

Subject to this approximation, it is easy to show that, when the
transmitting terminal is Alice, the test statistic $L$ is a
chi-square random variable with $2M$ degrees of freedom
\cite{Book:math}, i.e.,
\begin{align}
L=\frac{1}{\sigma ^2} \left ( \sum_{m=1}^M n_{rm}^2+\sum_{m=1}^M
n_{im}^2 \right ) \quad \sim \chi ^2 _{2M,0} ,
\end{align}
where $n_{rm}$ and $n_{im}$ are i.i.d Gaussian variables
$N(0,\sigma ^2)$. When the transmitting terminal is Eve, however,
$L$ becomes a non-central chi-square variable with a
non-centrality parameter $\mu _L$, i.e.,
\begin{align}
L&=\frac{1}{\sigma ^2} \left (\sum_{m=1}^M (\Delta
h^*_{rm}+n_{rm})^2+\sum_{m=1}^M
(\Delta h^*_{im}+n_{im})^2 \right ) \nonumber \\
&\sim \chi ^2_{2M,\mu _L} ,
\end{align}
where $\Delta h^*_{rm}$ and $\Delta h^*_{im}$ are the real and
imaginary part of ($H_{EBm}-{H}_{ABm} e^{j \phi^*}$),
respectively, with $\phi^*$ given by (\ref{l_phi}), and $\mu
_L=\frac{1}{\sigma ^2} \sum_{m=1}^M \mid {H}_{EBm}-{H}_{ABm} e^{j
\phi^*}  \mid^2$.

We define the rejection region for $H_0$ as $L>k$, where $k$ is
the threshold. Thus, the ``false alarm rate" (or Type I error) is

\begin{equation}
\alpha=P_{H_0}(L>k)=1-F_{\chi^2_{2M}}(k), \end{equation} and the
``miss rate" (or Type II error) is \begin{equation} \beta
=P_{H_1}(L<k)=F_{\chi^2_{2M,\mu_{L}}}(k), \end{equation} where
$F_X(\cdot)$ is the CDF of the random variable $X$. For a
specified $\alpha$, the threshold of the test is
$k=F^{-1}_{\chi^2_{2M}}(1- \alpha)$, and the miss rate is
$\beta=F_{\chi^2_{2M,\mu_{L}}}(F^{-1}_{\chi^2_{2M}}(1- \alpha))$,
which decreases with $\mu_{L}$. More specifically, with $\alpha$
fixed, $\beta$ rises with $\sigma ^2$ (because $k$ does) and falls
with $\sum_{m=1}^M \mid {H}_{EBm}-{H}_{ABm} e^{j \phi^*} \mid^2$.

\section{Simulation and Numerical Results}\label{sec_sim}
\subsection{Simulating the Transfer Functions}
In order to test the proposed scheme, it is necessary to model not
only ``typical" channel responses, but the spatial variability of
these responses. Only in that way can we discern the success in
detecting would-be intruders like Eve. To that end, we make use of
the WiSE tool, a ray-tracing software package developed by Bell
Laboratories \cite{WiSE:original}. One input to WiSE is the
3-dimensional plan of a specific building, including walls,
floors, ceilings and their material properties. With this
information, WiSE can predict the rays at any receiver from any
transmitter, including their amplitudes, phases and delays. From
this, it is straightforward to construct the transmit-receive
frequency response over any specified interval.

We have done this for one particular office building, for which a
top view of the first floor is shown in Fig. \ref{fig_building}.
This floor of this building is 120 meters long, 14 meters wide and
4 meters high. For our numerical experiment, we placed Bob in the
hallway (the filled-in circle) at a height of 2 m. For the
positions of Alice and Eve, we considered four rooms at the
extremities of the building (shown shaded). For each room, we
assumed Alice and Eve both transmitted from a height of 2 m, each
of them being anywhere on a uniform horizontal grid of points with
0.2-meter separations. With $N_s$ grid points in a room, there
were $N_s(N_s - 1)/2$ possible pairs of Alice-Eve positions. For
Rooms 1, 2, 3 and 4, the numbers of grid points were $N_s = 150$,
713, 315 and 348, respectively. For each Alice-Eve pair, (1) WiSE
was used to generate the Alice-Bob and Eve-Bob channel responses
($H_{AB}(f)$ and $H_{EB}(f)$); and (2) the hypothesis test
described above was used to compute $\beta$ for a specified
$\alpha$. The set of all $\beta$-values in a room were used to
compute a room-specific mean, $\overline{\beta}$, for each of
several selected combinations of bandwidth ($W$), number of tones
($M$) and transmit power ($P_T$).

\subsection{Transmit Power and Receiver Noise}
Assume that, in conjunction with WiSE, we obtain the various
transfer functions as dimensionless ratios (e.g., received
\emph{E}-field/transmitted \emph{E}-field). Then the proper
treatment of the noise variance, $\sigma^2$, in the hypothesis test
is to define it as the receiver noise power per tone, $P_N$, divided
by the transmit power per tone, $P_T/M$, where $P_T$ is the total
transmit power. Noting that $P_N  = \kappa TN_Fb$, where $\kappa T$
is the thermal noise density in mW/Hz, $N_F$ is the receiver noise
figure, and $b$ is the measurement noise bandwidth per tone in Hz,
we can write
\begin{align}
\sigma ^2 =\frac{\kappa TN_Fb}{P_T/M}=\frac{M}{\Gamma}
\end{align}
where $P_T$ is in mW, $\Gamma = P_T/P_N$  and we will henceforth
refer to $\Gamma$ by its decibel value.

\subsection{Simulation Results}

\begin{figure}[t]
\begin{center}
\epsfig{figure=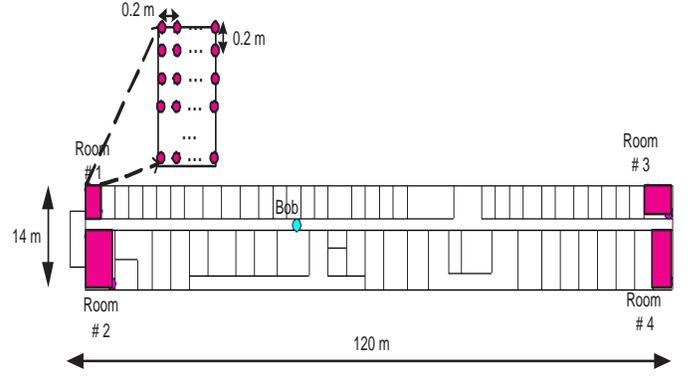, width=3.5in, height=2in}
\caption{System topology assumed in the simulations. Bob is located
at [45.6, 6.2, 3.0] m in a 120 m $\times$ 14 m $\times$ 4 m office
building. Alice and Eve are located on dense grids at a height of 2
m. The sizes of the grids are $N_s=150$, 713, 315, and 348,
respectively, for Room 1, 2, 3 and 4.}\label{fig_building}
\end{center}
\end{figure}

In the simulations, we set $\alpha=0.01$, $f_0=5$ GHz, $N_F=10$, and
$\Gamma=$ 90, 100, 110, 120 dB, which may be viewed as combinations
of $b=2.5$ MHz and $P_T=0.1$, 1, 10, 100 mW, respectively. As noted
earlier, we place Alice and Eve on dense grids in each of four rooms
at the corners of a particular building, with Bob in the hallway,
Fig. \ref{fig_building}.

We obtain a miss rate $\beta$ for each Alice-Eve pair, and then
calculate the mean value $\overline{\beta}$ for each room with
$M=1 \sim 10$ and $W=0.05 \sim 0.5$ GHz. The results verify the
utility of our algorithm and show that, if $P_T=100$ mW, most
values of $\overline{\beta}$ are below 0.05, even at the farthest
corners of the building.

\begin{figure}[t]
\begin{center}
\includegraphics[width=3.in]{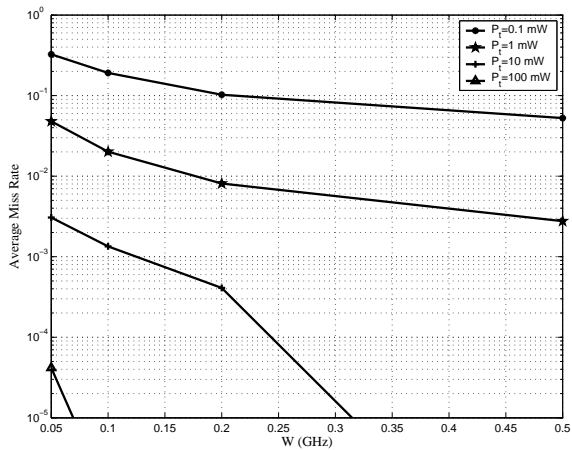} \\
(a) $M=5$ \\
\includegraphics[width=3.in]{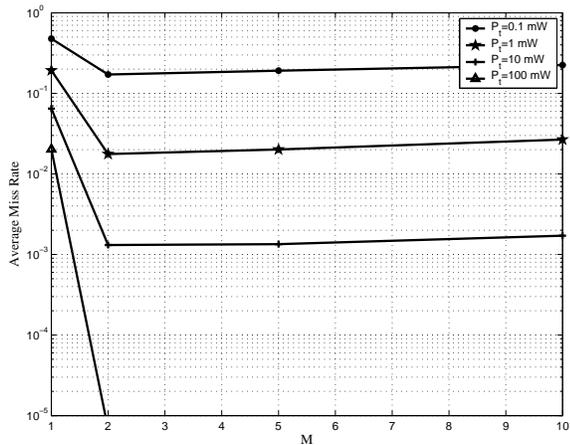} \\
(b) $W = 0.1$ GHz
\end{center}
\caption{Results for Room 1. Alice and Eve are placed within Room
1, while Bob is located in the center of the building, as depicted
in Fig. \ref{fig_building}. For each combination of Alice and Eve
locations, the corresponding channel responses to Bob were used to
estimate the miss rate. The average miss rate for Room 1,
$\overline{\beta}$, is reported as: (a) a function of bandwidth
($W$) for fixed number of tones ($M$); and (b) as a function of
$M$ for fixed $W$.}\label{F_r1}
\end{figure}

\begin{figure}[t]
\begin{center}
\includegraphics[width=3.in]{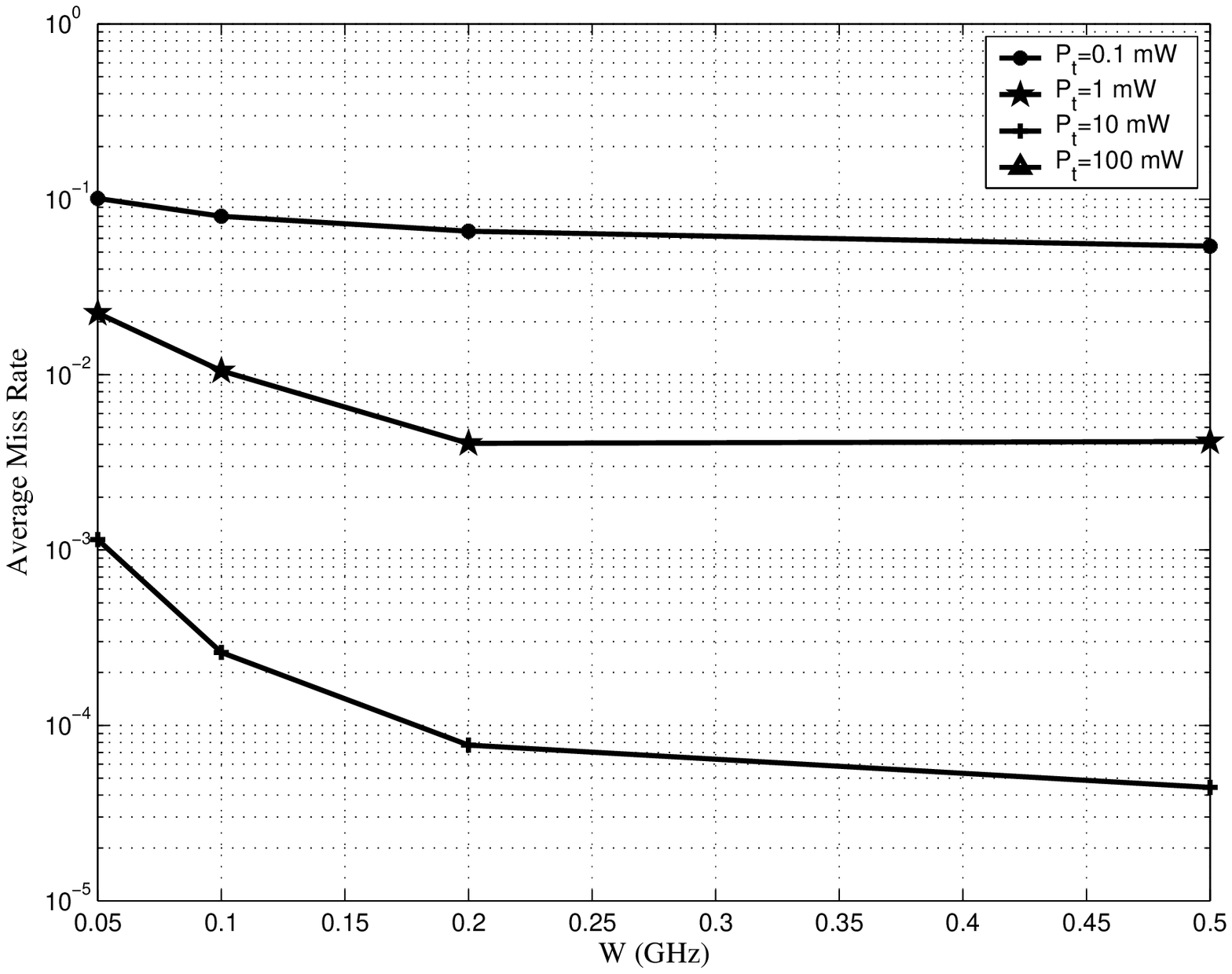} \\
(a) $M=5$ \\
\includegraphics[width=3.in]{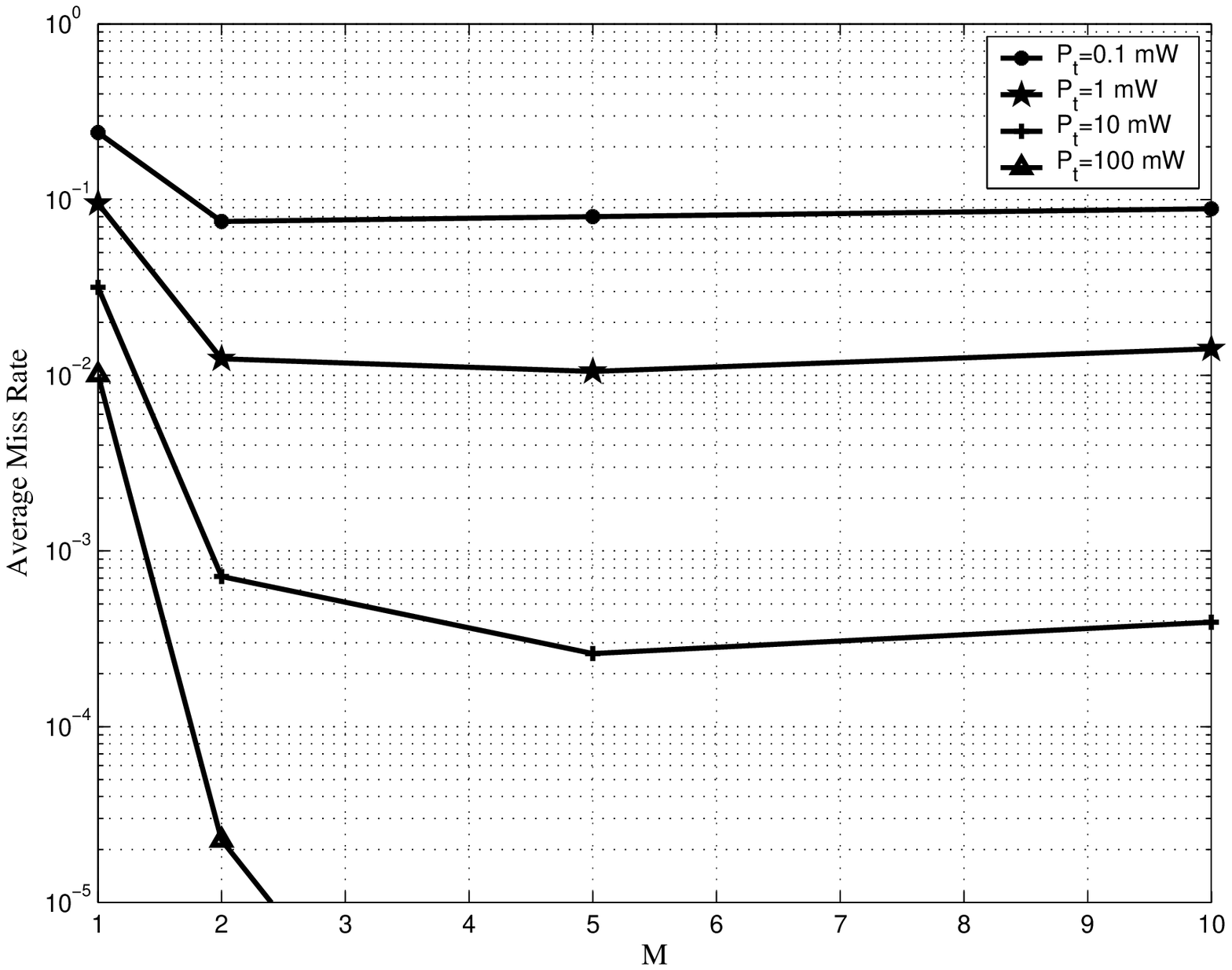} \\
(b) $W = 0.1$ GHz \end{center} \caption{The average miss rate,
$\overline{\beta}$, for Room 2, is reported as: (a) a function of
bandwidth ($W$) for fixed number of tones ($M$); and (b) as a
function of $M$ for fixed $W$.}\label{F_r2}
\end{figure}

\begin{figure}[t]
\begin{center}
\includegraphics[width=3.0in]{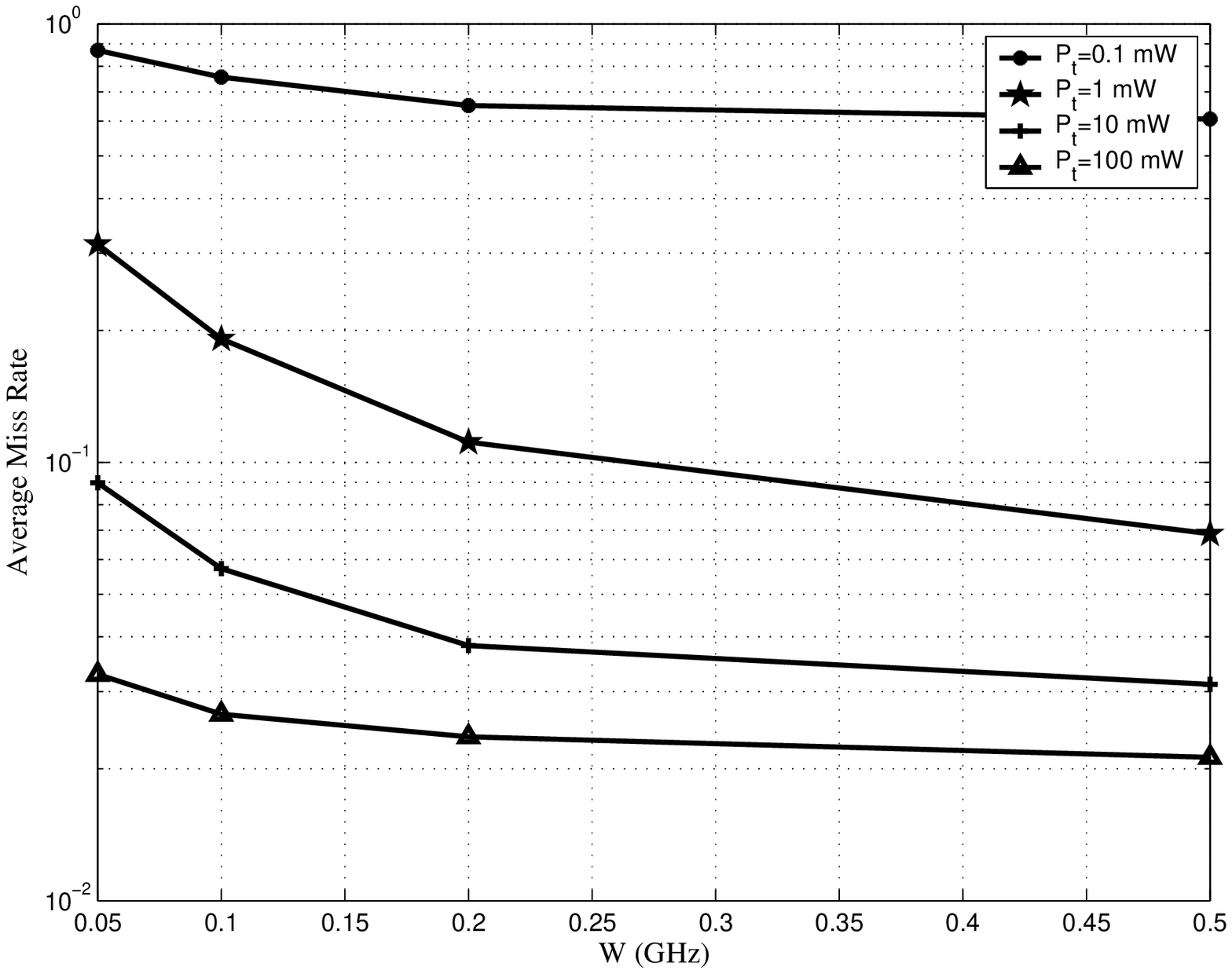} \\
(a) $M=5$ \\
\includegraphics[width=3.0in]{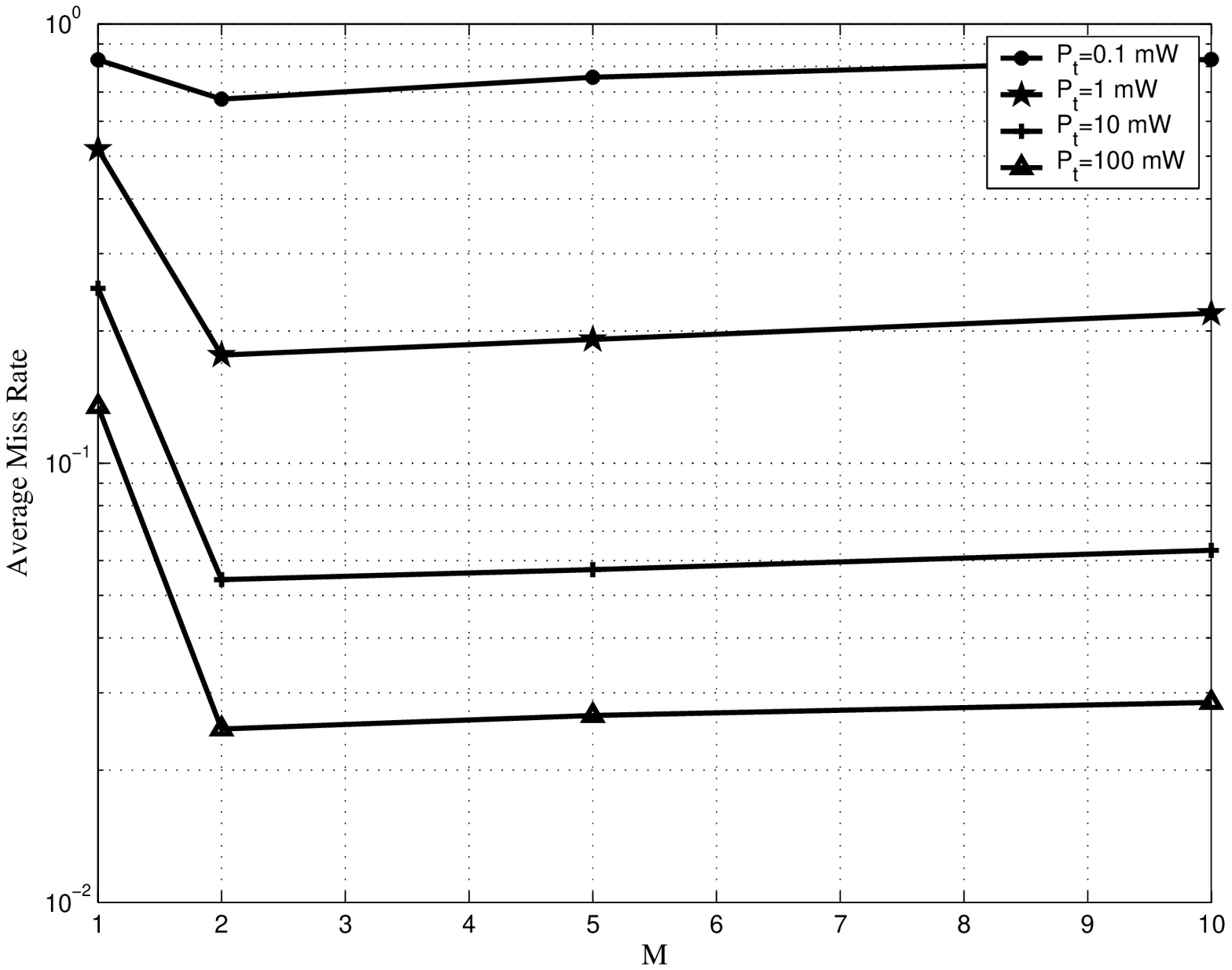} \\
(b) $W = 0.1$ GHz \end{center} \caption{The average miss rate,
$\overline{\beta}$, for Room 3, is reported as: (a) a function of
bandwidth ($W$) for fixed number of tones ($M$); and (b) as a
function of $M$ for fixed $W$.}\label{F_r3}
\end{figure}

\begin{figure}[t]
\begin{center}
\includegraphics[width=3.0in]{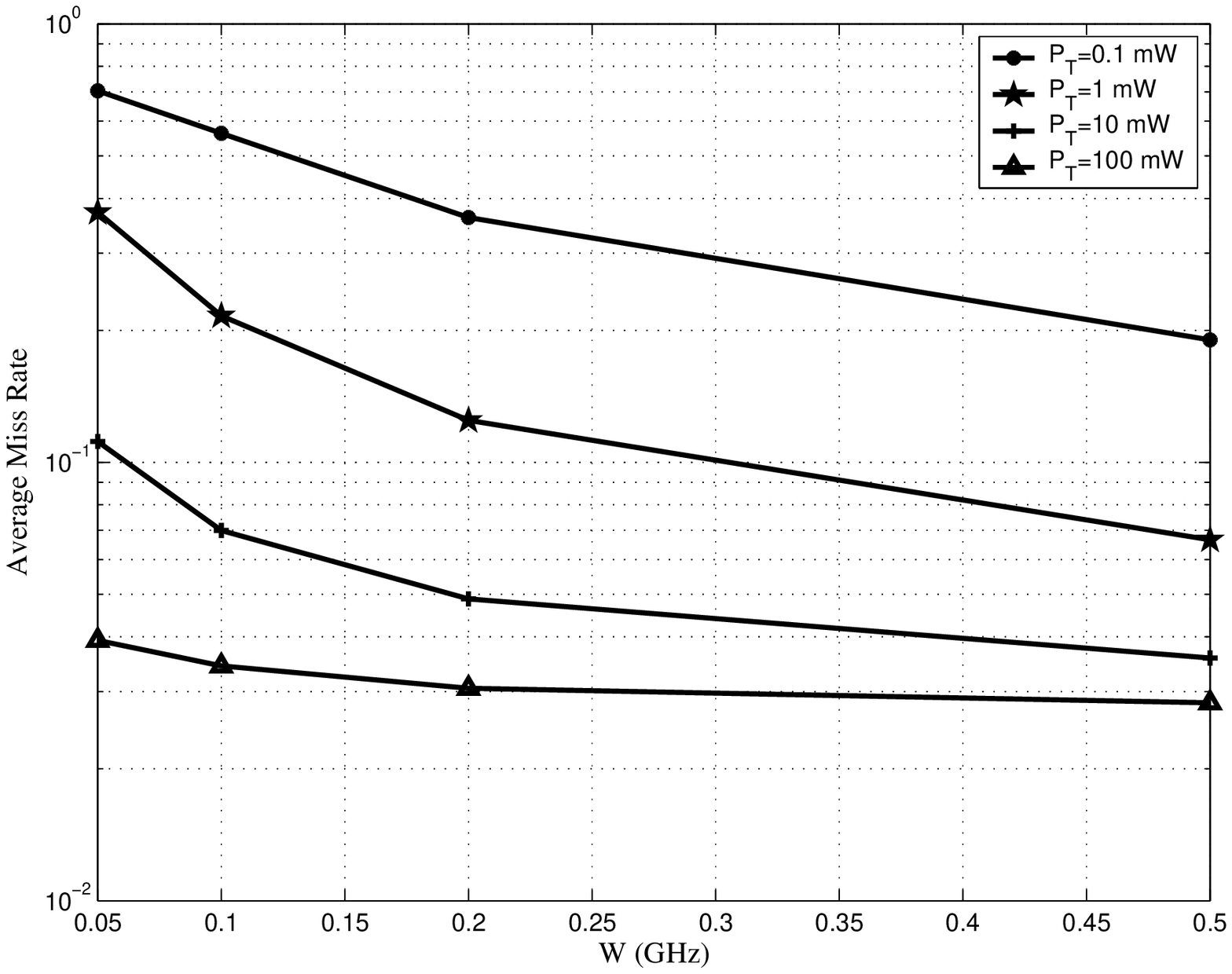} \\
(a) $M=5$ \\
\includegraphics[width=3.0in]{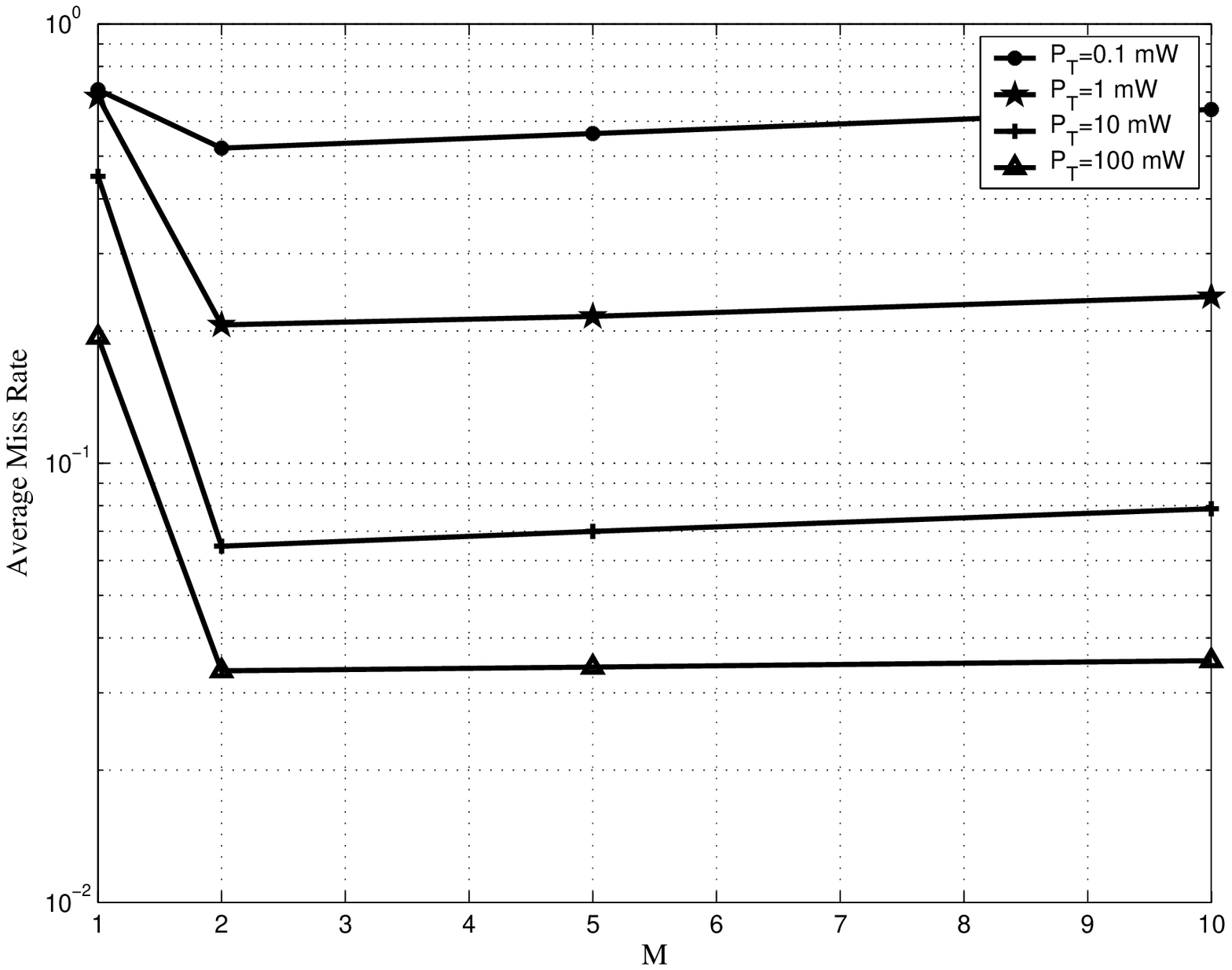} \\
(b) $W = 0.1$ GHz \end{center} \caption{The average miss rate,
$\overline{\beta}$, for Room 4, is reported as: (a) a function of
bandwidth ($W$) for fixed number of tones ($M$); and (b) as a
function of $M$ for fixed $W$.}\label{F_r4}
\end{figure}

Figures \ref{F_r1}-\ref{F_r4} show our computed results for Rooms
1-4, respectively. They show that, in terms of minimizing $\beta$,
increasing transmit power can be most beneficial, while increasing
the bandwidth and number of tones has less impact. In all cases,
there is little benefit (or even a deficit) in increasing $M$
beyond $\sim 5$; and in most cases, there is little benefit in
increasing $W$ beyond $ \sim 100-200$ MHz. This finding, however,
applies to the case where there are no temporal variations in the
levels or shapes of the transfer functions, a topic we discuss in
the last section.

Finally, the figures show the effects of distance (path length),
which influences the per-tone signal-to-noise ratios at Bob's
receiver. Rooms 3 and 4, which are farther from Bob than Rooms 1
and 2, have clearly poorer performance in rejecting Eve. Since the
four rooms are at the building extremities, we can assume that
this set of results lower-bounds the capabilities of our PHY-layer
authentication algorithm.

\section{Conclusion \& Future Work}\label{sec:Conclusion}
We have described and studied a physical layer technique for
enhancing authentication in a wireless in-building environment.
The technique uses channel frequency response measurements and
hypothesis testing to discriminate between a legitimate user
(Alice) and a would-be intruder (Eve). The study used the
ray-tracing tool WiSE to generate realistic spatially varied
responses, and results were obtained for several most-distant
(i.e., worst-case) rooms of one particular building. They confirm
the efficacy of the algorithm for realistic values of the
measurement bandwidth (e.g., $W \sim 100$ MHz), number of response
samples (e.g., $M \leq 5$) and transmit power (e.g., $P_T \sim
100$ mW). Computed results not shown here (but suggested by the
left side of Fig. 3a) indicate good performance down to $W=20$
MHz, so that the method can be used within bandwidths typical of
existing WLANs.

Moving forward, further investigation is needed to test other
buildings and to look at multiple Bob locations within the same
building, thereby establishing required power levels for a wider
class of cases. Another important topic is the temporal variations
of the measured channel responses, e.g., variations due to
movements within the building, slow time changes in the transmit
power and/or receiver noise level, etc. Our preliminary
investigations in \cite{L_SECU:var_tcomm} have confirmed the
efficacy of our approach in time-variant channels and showed that
the temporal variations even improve the performance in some
cases. Finally, as part of our ongoing efforts, we are working to
integrate physical layer authentication into a holistic
cross-layer framework for wireless security that will augment
traditional ``higher-layer" network security mechanisms with
physical layer methods.

%Moving forward, further investigation is needed to test other
%buildings and to look at multiple Bob locations within the same
%building, thereby establishing required power levels for a wider
%class of cases. Another important topic of further work is the
%temporal variations of the measured channel responses, e.g.,
%variations due to movements within the building, slow time changes
%in the transmit power and/or receiver noise level, etc. Where such
%changes are an issue, Bob's receiver must do more processing to
%ensure that Alice is not falsely rejected, which will have the
%effect of ``helping" Eve. We speculate that, in thus upgrading
%Bob's processing, we will identify a need for higher measurement
%bandwidths and/or more samples to obtain desired levels of
%performance. Finally, as part of our ongoing efforts, we are
%working to integrate physical layer authentication into a holistic
%cross-layer framework for wireless security that will augment
%traditional ``higher-layer" network security mechanisms with
%physical layer methods.

\end{document}